\documentclass[twocolumn,showpacs,preprintnumbers,amsmath,amssymb]{revtex4}

\usepackage{bm}

\newcommand{\bv}{\bar{\varphi}}

\begin{document}

\title{Quantum gravity and charge renormalization}

\author{David J. Toms}
\homepage{http://www.staff.ncl.ac.uk/d.j.toms}
\email{d.j.toms@newcastle.ac.uk}
\affiliation{%
School of Mathematics and Statistics,
Newcastle University,
Newcastle upon Tyne, U.K. NE1 7RU}

\date{\today}

\begin{abstract}
We study the question of the gauge dependence of the quantum gravity contribution to the running gauge coupling constant for electromagnetism. The calculations are performed using dimensional regularization in a manifestly gauge invariant and gauge condition independent formulation of the effective action. It is shown that there is no quantum gravity contribution to the running charge, and hence there is no alteration to asymptotic freedom at high energies as predicted by Robinson and Wilczek. 
\end{abstract}

\pacs{04.60.-m, 11.15.-q, 11.10.Gh, 11.10.Hi}

\maketitle

Until fairly recently the belief that quantum gravity lay outside the realm of any conceivable experimental test was pervasive in the physics community. Important work by Donoghue~\cite{Donoghue} helped to change this pessimistic viewpoint. Donoghue proposed that the effective field theory methodology be applied to quantum gravity, with Einstein's theory viewed as an effective low energy approximation to some as yet unknown more complete theory. Since Donoghue's pioneering work, there has been considerable interest in this viewpoint. A beautiful review of the subject has been given by Burgess~\cite{Burgess}.

A notable calculation was performed recently by Robinson and Wilczek~\cite{Wilczek} for Einstein gravity coupled to a gauge theory. It was claimed, as a consequence of quantum gravity corrections, that all gauge theories become asymptotically free at high energies. This includes the Einstein-Maxwell theory, and occurs below the Planck scale at which perturbative quantum gravity calculations become suspect. If the gravitational scale is sufficiently low, as predicted by many higher dimensional theories, then it is conceivable that the predictions of \cite{Wilczek} on the running gauge coupling constant might be experimentally testable~\cite{Gogoladze}. A recent analysis of the calculation~\cite{Wilczekcrit} has cast doubt on the results of \cite{Wilczek} by claiming that the quantum gravity correction to the running gauge coupling constant is gauge dependent, and that there is really no such effect. The calculations of \cite{Wilczekcrit} do demonstrate that when computed using traditional background-field methods the effective action does depend on the choice of gauge condition. Choosing a particular gauge condition, and demonstrating independence of parameters that enter this condition is not sufficient to demonstrate gauge condition independence. As we will discuss, with the choice of gauge conditions made in \cite{Wilczek,Wilczekcrit} a gauge condition independent result for the effective action necessitates additional terms that are not present if the standard background-field method is employed. Because of the interest in this problem, its potential as an experimental test of quantum gravity, and the controversy surrounding the details, we will present a different analysis to that of \cite{Wilczek,Wilczekcrit} that removes all question of any possible gauge dependence.

There are two  basic problems that need to be dealt with in any calculation in a gauge theory.  The first is to ensure that the calculations are done in a way that is gauge invariant.  The second is to ensure that the calculations are not dependent on any gauge conditions that are chosen.  The background-field method of DeWitt~\cite{DeWittdynamical} is the method that we will use here.  A version of the method that is invariant under gauge transformations of the background field was developed~\cite{DeWittQGII} and used by Abbott~\cite{Abbott} in Yang-Mills theory.  However, this formulation still leads to results that depend on gauge conditions in general.  A significant refinement of the method that leads to an effective action that is completely gauge-invariant as well as gauge condition independent was given by Vilkovisky~\cite{Vilk1} and DeWitt~\cite{DeWittVD}, and it is this formalism that we will use here.  (A recent review of the method, with applications, can be found in the forthcoming monograph~\cite{ParkerTomsbook}.) An alternate, but equivalent, approach is to use the conventional gauge-invariant background-field method, but to impose a set of identities as Nielsen~\cite{Nielsen} did for scalar electrodynamics to ensure that physical results do not depend on gauge conditions.  This approach has been used to great effect in gauge field theory at finite temperature~\cite{Kunstatteretal}.

The application of the Vilkovisky-DeWitt method to quantum gravity was used to solve the problem of the gauge condition dependence of the vacuum energy in five-dimensional Kaluza-Klein theory in Ref.~\cite{HKLT}.  Earlier applications to quantum gravity were given by \cite{BV,Fradkin}, and a more complete reference to other applications can be found in \cite{Odintsov,ParkerTomsbook}.

The central idea behind the Vilkovisky-DeWitt method is to regard the fields $\varphi^i$ as local coordinates in the space of all fields.  (We adopt DeWitt's condensed notation~\cite{DeWittdynamical} here.) The action functional $S\lbrack\varphi\rbrack$ is assumed to be invariant under the gauge transformations whose infinitesimal form is
\begin{equation}\label{eq1}
\delta\varphi^i=K^{i}_{\alpha}\lbrack\varphi\rbrack\delta\epsilon^\alpha\;,
\end{equation}
with $\delta\epsilon^\alpha$ the infinitesimal gauge parameters.  A metric $g_{ij}$ is introduced on the space of fields and an appropriate connection is chosen to ensure that the effective action is independent under arbitrary field redefinitions as well as independent under an arbitrary change of the gauge conditions.  (For proof of this and further details see \cite{EKT,ParkerTomsbook}.) Having established that the effective action is independent of the gauge condition, calculations are simplest if a special choice, called the Landau-DeWitt gauge condition by Fradkin and Tseytlin~\cite{Fradkin}, is made.  If we define the background field to be $\bv^i$ and write
\begin{equation}\label{eq2}
\varphi^i=\bv^i+\eta^i\;,
\end{equation}
the Landau-DeWitt gauge is specified by
\begin{equation}\label{eq3}
\chi_\alpha\lbrack\bv,\eta\rbrack=K_{\alpha i}\lbrack\bv\rbrack\eta^i=0\;.
\end{equation}
(Indices are raised and lowered with the field space metric in the usual way.) The calculational advantage of this choice is that the necessary connection $\Gamma^{k}_{ij}$ that is required may be taken as the usual Christoffel connection associated with the field space metric $g_{ij}$.  If any other gauge choice is made, this is no longer the case and additional terms in the connection beyond the Christoffel terms arise to ensure independence of the effective action on the gauge condition.  (An illustration of this for quantum gravity is given in \cite{HKLT}.)

To one-loop order the effective action is (see \cite{HKLT} for details of the derivation)
\begin{eqnarray}
\Gamma\lbrack\bv\rbrack&=&S\lbrack\bv\rbrack - {\rm ln\,det}\;Q_{\alpha\beta}\lbrack\bv\rbrack\nonumber\\
&&\hspace{-24pt}+\frac{1}{2}\lim_{\alpha\rightarrow0} {\rm ln\,det}\left\lbrace \nabla^i\nabla_j S\lbrack\bv\rbrack +\frac{1}{2\alpha}K^{i}_{\alpha}\lbrack\bv\rbrack K^{\alpha}_{j}\lbrack\bv\rbrack\right\rbrace\;, \label{eq4}
\end{eqnarray}
where
\begin{eqnarray}
\nabla_i\nabla_j S\lbrack\bv\rbrack&=&S_{,ij}\lbrack\bv\rbrack-\Gamma^{k}_{ij}\lbrack\bv\rbrack S_{,k}\lbrack\bv\rbrack\;,\label{eq5}\\
Q_{\alpha\beta}\lbrack\bv\rbrack&=&\frac{\delta\chi_\alpha}{\delta\epsilon^\beta}\;.\label{eq6}
\end{eqnarray}
It is worth emphasizing that the background field $\bv^i$ is completely arbitrary here, and it is not necessary to expand about classical solution to the equations of motion.  The role of non-trivial connection should be apparent when $\bv^i$ is general.

We are interested in the case of gravity with a Maxwell field here and will choose the curvature conventions of \cite{MTW} but with a Riemannian metric.  The classical action functional is
\begin{equation}\label{eq7}
S=\int d^nx|g(x)|^{1/2}\left(-\frac{2}{\kappa^2}R+\frac{1}{4}F_{\mu\nu}F^{\mu\nu}\right)
\end{equation}
where $\kappa^2=32\pi G$. The generic field variables will be chosen as $\varphi^i=(g_{\mu\nu}(x),A_\mu(x))$. As already noted, the formalism is completely independent of this choice, and we are free to choose $g^{\mu\nu}$ or $A^\mu$, or any arbitrary tensor densities without affecting the result for the effective action.  The infinitesimal gauge transformation (\ref{eq1}) for the gravity-Maxwell theory reads
\begin{eqnarray}
\delta g_{\mu\nu}&=&-\delta\epsilon^\lambda g_{\mu\nu,\lambda}-\delta\epsilon^{\lambda}{}_{,\mu}g_{\lambda\nu} -\delta\epsilon^{\lambda}{}_{,\nu}g_{\mu\lambda}\;,\label{eq10a}\\
\delta A_\mu&=&-\delta\epsilon^\nu A_{\mu,\nu}-\delta\epsilon^{\nu}{}_{,\mu}A_\nu +\delta\epsilon_{,\mu}\;,\label{eq10b}
\end{eqnarray}
in this case, where $\delta\epsilon^\mu$ describes an infinitesimal transformation of the spacetime coordinates, and $\delta\epsilon$ is the parameter for the local $U(1)$ gauge transformation of electromagnetism.  The natural choice of field space metric follows from the quadratic part of the expansion of the action functional about the background fields and is diagonal in the basic field variables with
\begin{equation}\label{eq11}
g_{g_{\mu\nu}(x)g_{\lambda\sigma}(x')}=\frac{1}{2}\left(g^{\mu\lambda}g^{\nu\sigma} +g^{\mu\sigma}g^{\nu\lambda}-g^{\mu\nu}g^{\lambda\sigma}\right)\delta(x,x')\;,
\end{equation}
the standard DeWitt metric for gravity, and
\begin{equation}\label{eq12}
g_{A_\mu(x)A_\nu(x')}=g^{\mu\nu}\delta(x,x')
\end{equation}
the field space metric for the Maxwell field.  Here $\delta(x,x')$ is the biscalar density Dirac distribution.  It can be verified that the $K^{i}_{\alpha}$ that can be read off by comparing (\ref{eq10a},\ref{eq10b}) with (\ref{eq1}) are Killing vectors for the field space metric.  It is a straightforward calculation to evaluate the Christoffel symbols for the field space metric given in (\ref{eq11},\ref{eq12}).

In order to check the results of \cite{Wilczek,Wilczekcrit} we do not need to evaluate the full one-loop effective action.  Instead we may note that we only need to study the terms of order $\bar{F}_{\mu\nu}\bar{F}^{\mu\nu}$, where $\bar{F}_{\mu\nu}=\bar{A}_{\nu,\mu}-\bar{A}_{\mu,\nu}$ is the field strength associated with the background electromagnetic gauge field $\bar{A}_\mu$.  The reason for this is that if we are interested in the corrections to quantum gravity in the $\beta$-function associated with the charge, standard renormalization group arguments in dimensional regularization \cite{tHooftRG} show that we can concentrate on pole terms in the renormalization factor $Z_e$ that links the bare charge $e_B$ to the renormalized charge $e_R$, $e_B=\ell^{n/2-2}Z_ee_R$ with $\ell$ an arbitrary unit of length.  The bare background field $\bar{A}_{\mu\,B}$ is related to the renormalized one $\bar{A}_{\mu\,R}$ by $\bar{A}_{\mu\,B}=\ell^{1-n/2}Z_A^{1/2}\bar{A}_{\mu\,R}$.  A consequence of using the gauge invariant background-field method is that $e_B\bar{A}_{\mu\,B}=e_R\bar{A}_{\mu\,R}$, leading to
\begin{equation}\label{eq13}
Z_eZ_A^{1/2}=1\;.
\end{equation}
This was used by Abbott~\cite{Abbott} in his calculation of the Yang-Mills $\beta$-function to two-loop order and is the background-field version of the Ward-Takahashi identity~\cite{Ward} in QED that shows how charge renormalization is determined by the renormalization of the photon field. Because the $\beta$-function for the charge is determined by the pole part of $Z_e$, and because $Z_e$ is related to $Z_A$ by (\ref{eq13}), we may concentrate on $Z_A$.  This is fixed by the pole part of the effective action $\Gamma$ that involves $\bar{F}_{\mu\nu}\bar{F}^{\mu\nu}$ as claimed.

Because we are only interested in checking the results of \cite{Wilczek,Wilczekcrit} we may concentrate on simply the pole part of $\Gamma$ that involves $\bar{F}_{\mu\nu}\bar{F}^{\mu\nu}$, rather than performing the more complicated calculation of the full one-loop effective action.  We can therefore set the background gravitational metric $\bar{g}_{\mu\nu}=\delta_{\mu\nu}$, but keep the background electromagnetic field $\bar{A}_\mu$ general.  This means that we are not expanding about a solution to the classical equations of motion, and that the contribution of the Vilkovisky-DeWitt correction to the traditional effective action is important.

A simple method to obtain all the terms of order $\bar{F}_{\mu\nu}\bar{F}^{\mu\nu}$ in $\Gamma$ is to reexpress the determinants in (\ref{eq4}) as functional integrals, and to treat the dependence on the background gauge field as an interaction.  In this way we can obtain a systematic expansion powers of $\bar{A}_\mu$, and in particular, concentrate on those expressions that are just quadratic in $\bar{A}_\mu$ since it is only these quadratic terms that can give rise to a dependence on $\bar{F}_{\mu\nu}\bar{F}^{\mu\nu}$ in $\Gamma$.

With the identification $\varphi^i=(g_{\mu\nu},A_\mu)$, and $\bar{\varphi}^i=(\delta_{\mu\nu},\bar{A}_\mu)$, it proves convenient to identify the quantum part of the field as
\begin{equation}\label{eq14}
\eta^i=(\kappa h_{\mu\nu},a_\mu)\;.
\end{equation}

The Landau-DeWitt gauge conditions (\ref{eq3}) read
\begin{eqnarray}
\chi_\lambda&=&\frac{2}{\kappa}(h_{\lambda}{}^{\mu}{}_{,\mu}-\frac{1}{2}h_{,\lambda})+\bar{A}_\lambda a^{\mu}{}_{,\mu}+a^\mu\bar{F}_{\mu\lambda}\;, \label{eq15a}\\
\chi&=&a^{\mu}{}_{,\mu}\;,\label{eq15b}
\end{eqnarray}
where $h=\delta^{\mu\nu}h_{\mu\nu}$ and spacetime indices are raised and lowered with the background metric $\delta_{\mu\nu}$.  Again we emphasize that any gauge choice can be made here, but that if anything other than the Landau-DeWitt gauge is chosen it is essential to include additional contributions to the connection, in addition to the Christoffel connection.

We can write the last term of (\ref{eq4}) as a functional integral over $h_{\mu\nu}$ and $a_\mu$ that is Gaussian.  The graviton and gauge field propagators are read off to be
\begin{eqnarray}
\langle h_{\alpha\beta}(x)h_{\lambda\tau}(x')\rangle&=&L_{\alpha\beta\lambda\tau}(x,x')\nonumber\\ &&\hspace{-24pt}=\int\frac{d^np}{(2\pi)^n}\,e^{ip\cdot(x-x')}\,L_{\alpha\beta\lambda\tau}(p)\;,\label{eq16}
\end{eqnarray}
where
\begin{eqnarray}
L_{\alpha\beta\lambda\tau}(p)&=&\frac{1}{p^2}\lbrack \delta_{\alpha(\lambda}\delta_{\tau)\beta} -\frac{1}{2(n-2)} \delta_{\alpha\beta}\delta_{\lambda\tau}\rbrack\nonumber\\
&&+\frac{(\alpha-1)}{p^4}\lbrack\delta_{\alpha(\lambda}p_{\tau)}p_{\beta} +\delta_{\beta(\lambda}p_{\tau)}p_{\alpha} \rbrack\;,\label{eq17}
\end{eqnarray}
and
\begin{eqnarray}
\langle a_{\mu}(x)a_{\nu}(x')\rangle&=&G_{\mu\nu}(x,x')\nonumber\\ &=&\int\frac{d^np}{(2\pi)^n}\,e^{ip\cdot(x-x')}\,G_{\mu\nu}(p)\;,\label{eq18}
\end{eqnarray}
where
\begin{equation}\label{eq19}
G_{\mu\nu}(p)=\frac{\delta_{\mu\nu}}{p^2}+(2\beta-1)\frac{p_\mu p_\nu}{p^4}\;.
\end{equation}
(The round brackets in (\ref{eq17}) denote a symmetrization over the enclosed indices with a factor of 1/2.)
Here we call the gauge parameters $\alpha$ and $\beta$ to distinguish those for the two gauge conditions (\ref{eq15a},\ref{eq15b}); both gauge parameters are taken to zero at the end of the calculation as required in the Landau-DeWitt gauge. Here $\langle\cdots\rangle$ denotes the evaluation of any expression using Wick's theorem with only one-particle irreducible graphs kept.

We can write
\begin{eqnarray}
-\frac{1}{2} {\rm ln\,det}\left\lbrace \nabla^i\nabla_j S\lbrack\bv\rbrack +\frac{1}{2\alpha}K^{i}_{\alpha}\lbrack\bv\rbrack K^{\alpha}_{j}\lbrack\bv\rbrack\right\rbrace&&\nonumber\\ &&\hspace{-108pt}=\ln\int\lbrack d\eta\rbrack\,e^{-S_q-S_{GF}}\;,\label{eq20}
\end{eqnarray}
where
\begin{equation}\label{eq21}
S_q=\frac{1}{2}\eta^i\eta^j\nabla_i\nabla_jS\lbrack\bv\rbrack=\frac{1}{2}\eta^i\eta^j\left(S_{,ij}-\frac{1}{2}\Gamma^{k}_{ij}S_{,k}\right)\;,
\end{equation}
\begin{eqnarray}
S_{GF}&=&\frac{1}{\kappa^2\alpha}\int d^nx\Big(h_{\nu}{}^{\mu}{}_{,\mu}-\frac{1}{2}h_{,\nu} +\frac{\kappa}{2}\bar{F}_{\mu\nu}a^\mu\Big)^2 \nonumber\\
&&+\frac{1}{4\beta}\int d^nx\left(a^{\mu}{}_{,\mu}\right)^2\;. \label{eq22}
\end{eqnarray}

The expression for $S_q$ is too lengthy to write out in full here, but is computed from the second functional derivative of the classical action in (\ref{eq7}) and the Christoffel connection for the field space metric evaluated at $g_{\mu\nu}=\delta_{\mu\nu}$ with $A_\mu=\bar{A}_\mu$ general.  The resulting expression has the same general structure as $S_{GF}$, being quadratic in the quantum fields $h_{\mu\nu}$ and $a_\mu$, and involving the background electromagnetic field $\bar{A}_\mu$ up to quadratic order.  The functional integral in (\ref{eq20}) is then computed perturbatively by writing $S_q+S_{GF}=S_0+S_1+S_2$ with the subscript counting powers of $\bar{A}_\mu$ that occur, and expanding the exponential in powers of $\bar{A}_\mu$ up to quadratic order.  We are interested in $\langle S_2\rangle$ and $\langle (S_1)^2\rangle$ to see if there are any poles that involve $\bar{F}_{\mu\nu}\bar{F}^{\mu\nu}$.  We find
\begin{eqnarray}
S_1&=&\frac{\kappa}{\alpha}\int d^nx(h^{\mu\nu}{}_{,\mu}-\frac{1}{2}h^{,\nu})a^\lambda\bar{F}_{\lambda\nu}\nonumber\\
&&\hspace{-24pt}+\frac{\kappa}{2}\int d^nx\Big(h\bar{F}^{\mu\nu}-2h^{\mu}{}_{\lambda}\bar{F}^{\lambda\nu}
+2h^{\nu}{}_{\lambda}\bar{F}^{\lambda\mu}\Big)a_{\nu,\mu}\label{eq23}
\end{eqnarray}
and a much lengthier expression for $S_2$.  ($S_0$ tells us the Feynman rules for the propagators given earlier in (\ref{eq16}--\ref{eq19}).) When computing $\langle S_2\rangle$ we encounter expressions like $\langle a_\mu(x)a_\nu(x)\rangle$ and $\langle h_{\mu\nu}(x)h_{\lambda\tau}(x)\rangle$ that involve the Feynman propagators with coincident arguments.  Such terms get regulated to zero in dimensional regularization~\cite{Leibbrandt}.  We conclude that $\langle S_2\rangle$ cannot contain pole terms proportional to $\bar{F}_{\mu\nu}\bar{F}^{\mu\nu}$.

The situation for $\langle (S_1)^2\rangle$ is less obvious and may give rise to pole terms in dimensional regularization; however, none of the poles can take the form of $\bar{F}_{\mu\nu}\bar{F}^{\mu\nu}$.  We will illustrate this for a typical term,
\begin{eqnarray}
&&\hspace{-8cm}\langle\partial_\mu h^{\mu\nu}(x)\partial_\alpha^\prime h^{\alpha\beta}(x')a^\lambda(x) a^\sigma(x') \rangle\nonumber\\
&&\hspace{-6cm}=\partial_\mu\partial_\alpha^\prime L^{\mu\nu\alpha\beta}(x,x')G^{\lambda\sigma}(x,x')\nonumber\\
=\int \frac{d^nk}{(2\pi)^n}e^{ik\cdot(x-x')}\Big\lbrace \frac{\alpha}{2}\delta^{\nu\beta} \int \frac{d^np}{(2\pi)^n} G^{\lambda\sigma}(k-p)&&\nonumber\\
+\Big(\frac{3\alpha}{2}-1-\frac{1}{(n-2)}\Big) \int \frac{d^np}{(2\pi)^n}
\frac{p^\beta p^\nu}{p^2}G^{\lambda\sigma}(k-p)\Big\rbrace&&\label{eq24}
\end{eqnarray}
The integral over $p$ in the first term is regulated to zero as described for $\langle S_2\rangle$.  The integral over $p$ in the second term does contain poles; however, these poles will involve powers of $k$ that result in derivatives of the Dirac $\delta$-distribution.  Such poles can give rise only to expressions involving higher derivatives, like $\bar{F}_{\mu\nu}\Box\bar{F}^{\mu\nu}$ and related terms, and are not of the required form.  It can be shown that this is true for all the terms that arise from $\langle (S_1)^2\rangle$.

The final term to check is ${\rm ln\,det}\;Q_{\alpha\beta}$.  This can be expressed as a functional integral by introducing Faddeev-Popov~\cite{Faddeev-Popov} ghosts.  In our case we need a vector ghost, $\eta^\mu$, and a scalar ghost, $\eta$, for the two gauge conditions (\ref{eq15a},\ref{eq15b}).  We can write
\begin{equation}\label{eq25}
{\rm det}\;Q_{\alpha\beta}=\int\lbrack d\bar{\eta}d\eta\rbrack\,e^{-S_{GH}}\;,
\end{equation}
where
\begin{eqnarray}
S_{GH}&=&\int d^nx\Big\lbrace -\frac{2}{\kappa^2}\bar{\eta}^\lambda\Box\eta_\lambda +\bar{A}_\lambda\bar{\eta}^\lambda\Box\eta -\bar{A}_\lambda\bar{A}_\nu\bar{\eta}^\lambda\Box\eta^\nu\nonumber\\
&&-\bar{\eta}^\lambda\bar{A}_{\lambda}(\bar{A}^{\mu}{}_{,\nu} +\bar{A}_{\nu}{}^{,\mu})\eta^{\nu}{}_{,\mu} -\bar{\eta}^{\lambda}\bar{A}_{\lambda}\eta^\nu\bar{A}_{\mu,\nu}{}^{\mu}\nonumber\\
&&-\bar{\eta}^{\lambda}\bar{F}^{\mu}{}_{\lambda}\eta^\nu\bar{A}_{\mu,\nu}-\bar{\eta}^{\lambda}\bar{F}^{\mu}{}_{\lambda}\bar{A}_\nu\eta^{\nu}{}_{,\mu}+\bar{\eta}^\lambda\bar{F}^{\mu}{}_{\lambda}\eta_{,\mu}\nonumber\\
&&+\bar{\eta}\Box\eta-\bar{\eta}\bar{A}_\nu\Box\eta^\nu -\bar{\eta}(\bar{A}^{\mu}{}_{,\nu}+\bar{A}_{\nu}{}^{,\mu})\eta^{\nu}{}_{,\mu}\nonumber\\
&&-\bar{\eta}\eta^\nu\bar{A}_{\mu,\nu}{}^{\mu}\Big\rbrace\;.\label{eq26}
\end{eqnarray}
This can be treated in the same way as we have just described, by expanding the exponential in powers of $\bar{A}_\mu$ up to quadratic order.  It is verified in a straightforward way that there are no poles that involve $\bar{F}_{\mu\nu}\bar{F}^{\mu\nu}$ coming from ${\rm ln\,det}\;Q_{\alpha\beta}$. (Again, higher derivative poles may occur.)

In conclusion, we have established in a gauge invariant and gauge condition independent way, in the framework of dimensional regularization, that there are no poles present in the one-loop effective action for the gravity-Maxwell theory that involve $\bar{F}_{\mu\nu}\bar{F}^{\mu\nu}$. This means that there is no infinite renormalization of the background gauge field that represents the photon, and hence no charge renormalization due to quantum gravity. We therefore disagree with the conclusions of \cite{Wilczek} and agree with those of \cite{Wilczekcrit}. Furthermore, unlike the analysis of \cite{Wilczekcrit}, this has been established in a way that is completely independent of any choice of gauge condition and for any arbitrary background gauge field.  Although we have only considered a $U(1)$ gauge theory here, sufficient evidence of the method has been presented to be confident that a similar conclusion holds for Yang-Mills theory coupled to gravity. It should be possible to compute the pole terms that involve derivatives and higher powers of $\bar{F}_{\mu\nu}$ within the scheme that we have described; this would represent a reflection of the non-renormalizability of the Einstein-Maxwell system first established by Deser and van Nieuwenhuizen~\cite{DeservanN}, but established in a manifestly gauge independent way.


\begin{thebibliography}{99}
\bibitem{Donoghue}
J.~F.~Donoghue, Phys. Rev. Lett. {\bf 72}, 2996 (1994); Phys. Rev. D {\bf 50}, 3874 (1994).

\bibitem{Burgess}
C.~P.~Burgess, Living Rev. Relativity {\bf 7}, 5 (2004); Online article: http://www.livingreviews.org/lrr-2004-5.

\bibitem{Wilczek}
S.~P.~Robinson and F.~Wilczek, Phys. Rev. Lett. {\bf 96}, 231601 (2006).

\bibitem{Gogoladze}
I.~Gogoladze and C.~N.~Cheung, Phys. Lett. B {\bf 645}, 451 (2007).

\bibitem{Wilczekcrit}
A.~R.~Pietrykowski, Phys. Rev. Lett. {\bf 98}, 061801 (2007).

\bibitem{DeWittdynamical}
B.~S.~DeWitt, {\it The Dynamical Theory of Groups and Fields\/} (Gordon and Breach, New York, 1965).

\bibitem{DeWittQGII}
B.~S.~DeWitt, in {\it Quantum Gravity II\/}, edited by C.~J. Isham, R. Penrose, and D.~W. Sciama (Oxford University Press, London, 1981).

\bibitem{Abbott}
L. Abbott, Nucl. Phys. B {\bf 185}, 189 (1981).

\bibitem{Vilk1}
G.~A. Vilkovisky, Nucl. Phys. B {\bf 234}, 125 (1984); in {\it The Quantum Theory of Gravity\/}, edited by S.~M. Christensen (Adam Hilger, Bristol, 1984).

\bibitem{DeWittVD}
B.~S. DeWitt, in {\it Quantum Field Theory and Quantum Statistics, Volume 1\/}, edited by I.~A. Batalin, C.~J. Isham, and G.~A. Vilkovisky (Adam Hilger, Bristol, 1987).

\bibitem{ParkerTomsbook}
L. Parker and D.~J. Toms, {\it Principles and Applications of Quantum Field Theory in Curved Spacetime\/} (Cambridge University Press, Cambridge, in press).

\bibitem{Nielsen}
N.~K. Nielsen, Nucl. Phys. B {\bf 101}, 173 (1975).

\bibitem{Kunstatteretal}
R. Kobes, G. Kunstatter, and A. Rebhan, Phys. Rev. Lett. {\bf 64}, 2992 (1990); Nucl. Phys. B {\bf 355}, 1 (1991).

\bibitem{HKLT}
S.~R. Huggins, G. Kunstatter, H.~P. Leivo and D.~J. Toms, Phys. Rev. Lett. {\bf 58}, 296 (1987); Nucl. Phys. B {\bf 301}, 627 (1988).

\bibitem{BV}
A.~O. Barvinsky and G.~A. Vilkovisky, Phys. Rep. {\bf 119}, 1 (1985).

\bibitem{Fradkin}
E.~S. Fradkin and A.~A. Tseytlin, Nucl. Phys. B {\bf 234}, 509 (1984).

\bibitem{Odintsov}
S.~D. Odintsov, Theor. Math. Phys. {\bf 82}, 45 (1990); Fortschr. Phys. {\bf 38}, 371 (1990).

\bibitem{EKT}
P. Ellicott, G. Kunstatter, and D.~J. Toms, Ann. Phys. (N.Y.) {\bf 205}, 70 (1991).

\bibitem{MTW}
C.~W. Misner, K.~S. Thorne and J.~A. Wheeler, {\it Gravitation\/} (W. H. Freeman, San Francisco, 1973).

\bibitem{tHooftRG}
G. `t~Hooft,  Nucl. Phys. B {\bf 61}, 455 (1973).

\bibitem{Ward}
J.~C. Ward, Phys. Rev. {\bf 78}, 182 (1950); Y. Takahashi, Nuovo Cimento {\bf 6}, 370 (1957); J. D. Bjorken and S.~D. Drell, {\it Relativistic Quantum Fields\/} (McGraw-Hill, New York, 1965).

\bibitem{Leibbrandt}
G. Leibbrandt, Rev. Mod. Phys. {\bf 47}, 849 (1975).

\bibitem{Faddeev-Popov}
L.~D. Faddeev and V.~N. Popov, Phys. Lett. B {\bf 25}, 29 (1967).

\bibitem{DeservanN}
S. Deser and P. van~Nieuwenhuizen, Phys. Rev. Lett. {\bf 32}, 245 (1974); Phys. Rev. D {\bf 10}, 401 (1974).
\end{thebibliography}
\end{document}